# Cosmological Parameters and Gravitational Lensing Statistics

PHILLIP HELBIG

*Hamburger Sternwarte · Gojenbergsweg 112*
*D-21029 Hamburg · Germany*

## INTRODUCTION

Lensing statistics gives a method of testing the cosmological constant at intermediate redshifts—at low and high redshifts the possibilities for measuring $\lambda_0$ are limited and even at intermediate redshifts $\lambda_0$ often cancels out in observable relations—using well-understood lensing theory and standard astrophysical assumptions.

It has recently been suggested by many authors (see, for example, Fukugita et al.[1] and references therein) that gravitational lensing statistics can provide a means of distinguishing between different cosmological models, most effectively concerning the value of the cosmological constant. Kochanek[2] has suggested a method based not on the total number of lens systems but rather on the redshift distribution of known lens systems characterised by observables such as redshift and image separation. Looking at a few different models, he concludes that flat, $\lambda$-dominated models are five to ten times less probable than more 'standard' models. The advantage of this method is that it is not plagued by normalisation difficulties as are most schemes involving the total number of lenses.

Since Kochanek was apparently able to get some interesting results using statistics based on only four gravitational lens systems, I wanted to exlpore this more thoroughly by looking at not just a few but all models characterised by $\lambda_0$ and $\Omega_0$ as well as varying degrees of homogeneity. I also looked at selection effects and did some simulations to get a handle on what the results mean.

## THEORY

I make the 'standard assumptions' that the Universe can be described by the Robertson-Walker metric and that lens galaxies can be modelled as non-evolving singular isothermal spheres (SIS). This leads an equation for the relative differential optical depth[3]

$$\frac{\mathrm{d}\tau}{\mathrm{d}z_\mathrm{d}} = (1+z_\mathrm{d})^2 \frac{a}{a*}\frac{\gamma}{2}\left(\frac{a}{a*}\frac{D_\mathrm{s}}{D_\mathrm{ds}}\right)^{\frac{\gamma}{2}(1+\alpha)} \qquad \times$$

$$D_{\mathrm{d}}^2 \frac{1}{\sqrt{Q(z_{\mathrm{d}})}} \exp\left(-\left(\frac{a}{a*}\frac{D_{\mathrm{s}}}{D_{\mathrm{ds}}}\right)^{\frac{\gamma}{2}}\right) \tag{1}$$

where $a* := 4\pi\left(\frac{v*}{c}\right)$ ($v* := v$ of an $L*$ galaxy), $\gamma$ is the Faber-Jackson/Tully-Fisher exponent, $\alpha$ the Schechter exponent, $D_{\mathrm{d}}$ the angular size distance between the observer and the lens and

$$Q(z_{\mathrm{d}}) := (1+z_{\mathrm{d}})^2(\Omega_0 z_{\mathrm{d}} + 1 - \lambda_0) + \lambda_0. \tag{2}$$

The optical depth depends on the cosmological model through $Q(z_{\mathrm{d}})$ as well as through the angular size distances, because of the fact that $D_{ij} = D_{ij}(z_i, z_j; \lambda_0, \Omega_0, \eta)$. The influence of $\eta$, which gives the fraction of homogeneously distributed, as opposed to compact, matter is felt only in the calculation of the angular size distances, whereas the cosmological model in the narrower sense makes its influence felt here as well as through $Q(z_{\mathrm{d}})$. The angular size distances can be calculated for arbitrary cosmological models by the procedure given in Kayser & Helbig.[4]

## CALCULATIONS

The following gravitational lens systems meet my selection criteria: 0142-100 (UM 673), 0218+357, 1115+080 (Triple Quasar), 1131+0456, 1654+1346 and 3C324. (For more details on these systems see Refsdal & Surdej.[5])

I considered the following ranges of values for the cosmological parameters:

$$\begin{array}{rcccl} -10 & < & \lambda_0 & < & +10 \\ 0 & < & \Omega_0 & < & 10 \end{array}$$

In order to measure the relative probability of a given cosmological model, I defined the quantity $f$ as follows: p

$$0 < f := \frac{\int_0^{z_{\mathrm{l}}} \mathrm{d}\tau}{\int_0^{z_{\mathrm{s}}} \mathrm{d}\tau} < 1, \tag{3}$$

where $z_{\mathrm{l}}$ is the *observed* lens redshift for a particular system. ($z_{\mathrm{d}}$ is used to denote the variable corresponding to lens redshift as opposed to the measured value for a particular lens.) The distribution of the different $f$ values (one for each lens system in the sample) in $b$ equally-sized bins in the interval ]0,1[ gives the relative probability $p$ of a given cosmological model, with

$$p = \prod_{i=1}^{b} \frac{1}{n_i!} \tag{4}$$

where $n_i$ is the number of systems in the $i$-th bin. (This definition allows only a few discrete values, of course.) The variable $b$ is a free parameter; since the most information is obtained when $b$ is equal to the number of systems, I adopt this value for $b$.

## RESULTS AND DISCUSSION

My results are in Fig. 1, plot **b**. (This is for $\eta = 0.5$; the results do not depend strongly on $\eta$.[3] See Kayser & Helbig[4] for a discussion of this parameter.) One can see basically that areas of equal probability occur in some fashion which is not stochastic. Although there are only a few discreet values for the probability as defined in Eq. (4), nevertheless one sees a degeneracy—there is a wide range of cosmological models for a given probability. Plot **c** shows the result of neglecting the observational bias, *e.g.* assuming that the lens could have its redshift measured whatever this redshift were. As a comparison with plot **b** shows, this leads to a bias against models with a high median expected lens redshift—those near the de Sitter model.

For comparison, I have also tested the method on the systems used by Kochanek,[2] using $m_{\mathrm{lim}} = \infty$ und $\eta = 1$, both of which he implicitly assumes. (Of course, when one considers finite values for $m_{\mathrm{lim}}$, one cannot include systems with lens redshifts which have been determined by means other than measured emission redshifts, such as absorption lines (which assumes that the lens is also the absorber).) The results are in plot **d** where the relative probabilities are 0, $\frac{1}{6}$, $\frac{1}{2}$ and 1 and comparing the various models examined by Kochanek confirm his conclusions. For instance, the relative probabilities of the Einstein-de Sitter and de Sitter model are 1 and $\frac{1}{6}$, confirming his result that flat, $\lambda$-dominated models are 5–10 times less probable than standard ones. (However, taking $m_{\mathrm{lim}}$ into account and/or using only directly measured lens redshifts would produce quite different results, as discussed above.) This plot artificially indicates a low probability for models near the de Sitter model for the same reasons as those discussed in connection with plot **c**.

## NUMERICAL SIMULATIONS

For the numerical simulations, the observables $\theta''$ (the radius of the Einstein ring or *half* the image separation corresponding to the diameter of the Einstein ring), $z_{\mathrm{s}}$ and galaxy type were chosen randomly from an interval roughly corresponding to the observed range of values in order to produce synthetic data comparable to real observations. For a given cosmological model, the corresponding lens redshift $z_{\mathrm{l}}$ for each system was calculated from the observables and a randomly generated $f$ through (numerical) inversion of Eq. (3). This catalog was then used to determine a relative probability for each of the points in the $\lambda_0$-$\Omega_0$ plane in the same manner as for the real systems.

The Kolmogorov-Smirnov (K-S) test is of course a well understood method for testing if two distributions are statistically significantly different. (See, e.g., Press et.al.[6] for a general discussion and definition of the K-S probability.) However, this test can only be used for distributions with more than $\approx 20$ data points. Therefore I plot in Fig. 1 in plots **b**, **c**, and **d** the probability given by Eq. (4) and in plot **e** the K-S probability.

I have done simulations for a variety of world models and also for numbers of systems between 20 and 50. In the interest of brevity, I present only one plot. Plot **e** in Fig. 1 shows the results derived from a catalogue of simulated gravitational lens systems. Since, even with 50 systems, no area can be excluded based on the K-S

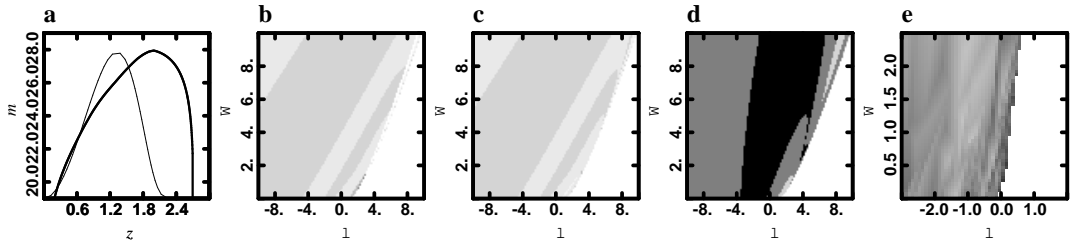

Figure 1: **a.** The relative differential optical depth (thin curve) and the calculated lens brightness $m$ (thick curve) as functions of $z_d$. The world model is the de Sitter model ($\lambda_0 = 1.0, \Omega_0 = 0.0$) and the observables are those for the gravitational lens system $0142-100$. The ordinate gives the magnitude in Johnson $R$. For realistic limiting spectroscopic magnitudes ($\approx 24^m$) it is clear that one cannot sample the probability distribution without a strong bias. **b.** Relative probability for the systems mentioned in the text, with a realistic value for $m_{\lim}$. Here and in the other halftone plots, the relative probability increases linearly from white to black. **c.** The same as b., but *neglecting* $m_{\lim}$, which, as in plot d., makes the models near the de Sitter model appear more probable than they are. **d.** Relative probability for the systems used by Kochanek. **e.** Results based on a catalogue of 50 simulated systems. (Note the different scale on the axes.) The cosmological model used to generate the lens redshifts is the homogeneous Einstein-de Sitter model ($\eta = 1.0, \lambda_0 = 0.0, \Omega_0 = 1.0$).

probability—the white area has $p = 0$ due to the fact that at least one lens would be fainter than $m_{\lim}$ in these world models, as discussed in Helbig & Kayser[3]— I conclude that, although one can qualitatively understand the physics which at least in part is responsible for the results presented in Fig. 1, the actual relative probabilities are more indicative of intrinsic scatter in the redshifts of the lenses than a hint of the correct cosmological model.

## SUMMARY AND CONCLUSIONS

For known gravitational lens systems the redshift distribution of the lenses was compared with theoretical expectations for $10^4$ Friedmann-Lemaître cosmological models, which more than cover the range of possible cases. The comparison was used for assigning a relative probability to each of the models. However, my simulations indicate that a reasonable number of observed systems cannot deliver interesting constraints on the cosmological parameters using this method. Therefore, it seems that lensing statistics can tell us something about the cosmological model only if one makes use of all information, which means coming to grips with normalisation difficulties.